\documentclass[12pt]{iopart}
\usepackage{mathrsfs}
\usepackage{graphicx}
\usepackage{dcolumn}
\usepackage{bm}

\begin{document}
\title[Dark-Sates in EIT
Controlled by a Microwave Field]{Dark-Sates in Electromagnetically
Induced Transparency Controlled by a Microwave Field}
\author{Bin Luo$^1$, Hua Tang$^2$ and
Hong Guo$^1$}
\address{$^1$ CREAM Group, State Key Laboratory of Advanced
Optical Communication Systems and Networks (Peking University) and
Institute of Quantum Electronics, School of Electronics Engineering
and Computer Science, Peking University, Beijing 100871, People's
Republic of China}
\address{$^2$ School of Electronics and Information Engineering, Beihang
University, Beijing 100191, People's Republic of China}
\ead{hongguo@pku.edu.cn}
\begin{abstract}
The rigorous dark-sate conditions, i.e., two-photon resonance, for a
$\Lambda$-type electromagnetically induced transparency (EIT) are
extended with the system controlled by a microwave field. The
extended dark-states are found to be extremely sensitive to the
phase of the microwave field in a narrow interval. However, the
dark-state is no longer the requirement for EIT in this scheme and
EIT without dark states is present.
\end{abstract}
\pacs{42.50.-p; 42.50.Hz; 42.50.Md} \submitto{\jpb} \maketitle
\section{Introduction}

Electromagnetically induced transparency has drawn tremendous
attention during the last two decades \cite{Harris, RMPMF}. The
transparency together with dramatic dispersion has wide applications
in light velocity control \cite{Hau, Akulshin1998}, nonlinear optics
\cite{hemmer1995, MXiao1, scully1999, harris2004, Mxiao} and quantum
information \cite{MFandL, MFandL2, Bergman, DLCZ, Lukinns, kuzmich}.
Dark-states, as the key property of EIT, are of great importance in
quantum control of atom and optical quanta \cite{MFandL, MFandL2}.
The existence of dark-states are significant to an EIT based medium,
especially for optical processing and quantum manipulations such as
coherent population trapping (CPT), coherent population transfer
\cite{Bergman} and photon storage \cite{MFandL}.

Microwaves coupling the ground states can manipulate the
characteristics of the EIT medium
\cite{DV,wilson,shahriar,gemma,sun,agarwal,arx}. For a closed
system, the population trapping is very sensitive to the relative
phase of the electromagnetic fields \cite{DV}. Microwave interaction
has been applied to excite the Raman trapped state and to influence
the CPT in a $\Lambda$ system \cite{shahriar}. Fast and slow light
phenomena in EIT medium controlled by microwave have been
theoretically predicted \cite{sun,agarwal}. Constructive and
destructive interference in the presence of microwave field has been
experimentally realized based on V-type system in solid
($\mathrm{Pr^{3+}YAlO_3}$) \cite{gemma} and recently in
$\Lambda$-type atom vapor \cite{arx}.

In this paper, we investigate a $\Lambda$-type system controlled by
a microwave field coupling ground states as in \cite{shahriar, sun,
arx}. Two-photon detuning condition has been extended and the phase
sensitivity of the dark-states is studied. With significant ground
states dephasing the dark states are proved to be vanished, while
the transparency still exists provided that the intensity and phase
of the microwave field are chosen appropriately. That is to say, EIT
can be established in severe dephasing medium without dark-states.

\section{Rigorous Dark-Sate Condition}
\begin{figure}
\centering
  \includegraphics[width=6 cm]{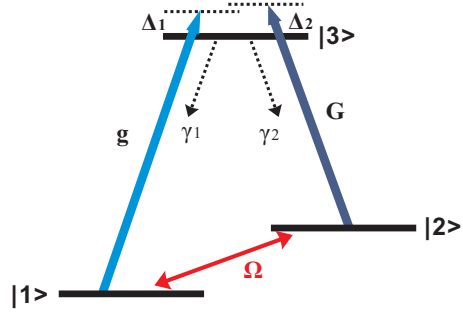}\\
  \caption{Interaction scheme. A $\Lambda$-type EIT medium is controlled by a microwave field coupling the ground states.
  Each electromagnetic field is labeled by its own Rabi frequency. }\label{levels}
\end{figure}

Consider a closed three-level $\Lambda$-type system shown in
\Fref{levels}. Basically, it is a $\Lambda$-type EIT system with
probe and pumping optical fields coupling the ground states
$|1\rangle$ and $|2\rangle$ to the excited state $|3\rangle$. A
controlling microwave fields with relative phase $\phi$ to the
optical fields is applied between the two grounds states. Under

\begin{eqnarray}\label{Hamiltonian}
    H&=&\hbar \Delta_1|1\rangle\langle1|+\hbar
    \Delta_2|2\rangle\langle2|\nonumber\\&&-\hbar\left(g|1\rangle\langle3|+G|2\rangle\langle3|
    +\Omega \rme^{\rmi\phi}|1\rangle\langle2|+\mathrm{h.c.}\right),
\end{eqnarray}
where $g$, $G$ and $\Omega$ are Rabi frequencies of the probe light,
pumping light and microwave field, respectively. $\Delta_1$ and
$\Delta_2$ are optical detunings of the probe and pumping, and the
microwave field is assumed to be resonant for simplicity.

The eigenstates of the system are
\begin{eqnarray}\label{eigenstates}
    |\psi_i\rangle&=&-\rme^{\rmi\phi}[G^2+\lambda_i(\Delta_2-\lambda_i)]|1\rangle+
    (\rme^{\rmi\phi}gG-\Omega\lambda_i)|2\rangle\nonumber\\&&+[G\Omega+\rme^{\rmi\phi}g(\Delta_2-\lambda_i)]|3\rangle,
\end{eqnarray}
where $\lambda_i=\lambda_{1,2,3}$ are eigenvalues of the Hamiltonian
and satisfy
\begin{eqnarray}\label{characteristicequation}
    &&\lambda^3-\lambda^2(g^2+G^2+\Omega^2)+2gG\Omega
    \cos{\phi}\nonumber\\&&+(g^2-\lambda^2)\Delta_2+(G^2-\lambda^2)\Delta_1+\lambda\Delta_1\Delta_2=0.
\end{eqnarray}

A dark-state requires elimination of the excited state $|3\rangle$,
which means $G\Omega+\rme^{\rmi\phi}g(\Delta_2-\lambda_i)=0$. Since
the eigenvalue of the Hamiltonian is always real, $\rme^{\rmi\phi}$
needs to be real and so $\phi=0$ or $\pi$.

(a) If $\phi=0$, one has $G\Omega+g(\Delta_2-\lambda_i)=0$.
Substituting it into \eref{characteristicequation} yields
\begin{equation}\label{c1}
    \Delta_1-\Delta_2=-\frac{\Omega}{gG}(g^2-G^2),
\end{equation}
where $\Delta_1-\Delta_2$ is called the two-photon detuning. This is
the dark-state condition in our scheme. In absence of the magnetic
field, $\Omega=0$, this condition can be reduced into two-photon
resonance $\Delta_1=\Delta_2$. Dressed states under this condition
are

\begin{eqnarray}
  |\lambda^0_0\rangle &:\;& G|1\rangle-g|2\rangle, \nonumber\\
  |\lambda^0_\pm\rangle &:\;&
  g|1\rangle+G|2\rangle+\lambda_\mp|3\rangle,
\end{eqnarray}
where $\lambda^0_{0,\pm}$ are eigenvalues with $\phi=0$,
\begin{eqnarray}
  \lambda^0_0 &=& \frac{g \Omega }{G}+\Delta _1, \nonumber\\
  \lambda^0_\pm &=& \frac{-G \Omega +g \Delta _1\pm\sqrt{4 \left(g^2+G^2\right) g^2+\left(G \Omega -g
   \Delta _1\right){}^2}}{2 g}.\nonumber
\end{eqnarray}

(b) If $\phi=\pi$, following the same process, the dark state
condition is derived as
\begin{equation}\label{c2}
    \Delta_1-\Delta_2=\frac{\Omega}{gG}(g^2-G^2),
\end{equation}
with eigenvalues and eigenstates as
\begin{eqnarray}
  |\lambda^\pi_0\rangle &:\;& G|1\rangle-g|2\rangle, \nonumber\\
  |\lambda^\pi_\pm\rangle &:\;&
  g|1\rangle+G|2\rangle+\lambda_\mp|3\rangle,
\end{eqnarray}
where
\begin{eqnarray}
  \lambda^\pi_0 &=& \frac{g \Omega }{G}+\Delta _1, \nonumber\\
  \lambda^\pi_\pm &=& \frac{G \Omega +g \Delta _1\pm\sqrt{4 \left(g^2+G^2\right) g^2+\left(G \Omega -g
   \Delta _1\right){}^2}}{2 g}.\nonumber
\end{eqnarray}

For both above cases without microwave, the corresponding states
reduce to the same dark-state as in EIT under the same two-photon
resonance condition. With a microwave field, however, the two-photon
resonance condition can be extended. A direct result shows that the
two-photon detuning can vary in a wide range by only changing the
intensity of the microwave field.

Dark-states in our scheme are the result of the coherence among the
two optical fields and the microwave field. Without the microwave
field, the dark state, as in EIT, is
$|D\rangle=(G|1\rangle-g|2\rangle)/\sqrt{g^2+G^2}$. In comparison,
the dressed states of the microwave transitions are
$(\rme^{\rmi\phi}|1\rangle \pm|2\rangle)/\sqrt{2}$. If no coherence
happens \cite{shahriar}, the direct transition of the dark-state to
the microwave dressed states requires $\phi=0$ or $\pi$, $g=G$ and
$\Delta_1=\Delta_2$, which is only a special case of the condition
\eref{c1} and \eref{c2}. However, it is the coherent interaction
that promotes the two-photon resonance condition into the condition
that the two-photon detuning is controllable. For the microwave
field, the dark-state has changed the population distribution of the
components in the dressed states, which is not possible in the
absence of the optical fields. The coherent process benefits both
the optical EIT dark-states and the microwave fields.

\section{Dampings included: the Density Operator Approach}
It is well known that the most important property of dark-state is
the elimination of the excited state and hence the spontaneous
emission can be ignored. Since the dephasing between ground states
are usually negligible for atom vapor, dark-states could remain very
stable in such medium. However, in some cases when the ground states
dephasing is significant, the dark-state condition has to be
justified and the influence of the dephasing has to be considered.
For this reason the density operator approach is required. For the
Hamiltonian shown in \eref{Hamiltonian}, the matrix equations reads:

\begin{eqnarray}\label{densitymatrix}
    \dot{\rho}_{11}&=&2\gamma_1\rho_{33}+\left(\rmi g \rho_{31}+ \rmi \rho_{21}\Omega \rme^{\rmi
    \phi}+\mathrm{h.c.}\right),\nonumber\\
    \dot{\rho}_{22}&=&2\gamma_1\rho_{33}+\left(\rmi G \rho_{32}- \rmi \rho_{21}\Omega \rme^{\rmi
    \phi}+\mathrm{h.c.}\right),\nonumber\\
    \dot{\rho}_{21}&=&\left[\rmi(\Delta_1-\Delta_2)-\kappa\right]\rho_{21}-\rmi g\rho_{23}+\rmi G\rho_{31}\nonumber\\
    &&+i\Omega \rme^{-\rmi\phi}(\rho_{11}-\rho_{22}),\nonumber\\
    \dot{\rho}_{31}&=&\left[\rmi\Delta_1-(\gamma_1+\gamma_2)\right]\rho_{31}-\rmi\rho_{32}\Omega
    \rme^{-\rmi\phi}+\rmi G\rho_{21}\nonumber\\&&+\rmi
    g(\rho_{11}-\rho_{33}),\nonumber\\
    \dot{\rho}_{32}&=&\left[\rmi\Delta_2-(\gamma_1+\gamma_2)\right]\rho_{32}-\rmi\rho_{31}\Omega
    \rme^{\rmi\phi}+\rmi g\rho_{12}\nonumber\\&&+\rmi
    g(\rho_{22}-\rho_{33}),\nonumber
\end{eqnarray}
where $\gamma_{1,2}$ are the radiation damping rates from the
excited states to $|1\rangle$ and $|2\rangle$, respectively.
$\kappa$ is the dephasing rate between ground states.

The dark states condition requires $\rho_{33}$=0. On resonance case
($\Delta_1=\Delta_2=0$), the dark-state condition is reduced to
\begin{eqnarray}\label{rhodk}
\cos^2{\phi}=1+\frac{2\gamma}{g^2G^2\Omega^2(4\gamma+\kappa)^2}
\left\{2g^2G^2\gamma\kappa^2+\right.\nonumber\\
\left.4\kappa\Omega^2\gamma^2\left(g^2+G^2\right)+\left(g^2-G^2\right)^2\left[\kappa
(g^2+G^2)+2\gamma\Omega^2\right]\right\}.
\end{eqnarray}
Only when $g=G$ and $\kappa=0$ can $\phi$ be real with
$\cos{\phi}=\pm 1$. This agrees exactly with the dark-state
condition \eref{c1} and \eref{c2} on resonance case. For
non-resonant case, the dark-state condition is the same as \eref{c1}
and \eref{c2} with $\kappa=0$ (see appendix). \Fref{phi332} shows
that with different $\phi$, the excited state population changes
from high occupation to zero. It is obvious that the existence of
the dephasing term will inevitably cause $\cos{\phi}>1$ whether the
microwave field is applied or not, and so that the dark-states can
never be achieved.

\begin{figure}
\centering
  \includegraphics[width=6.5 cm]{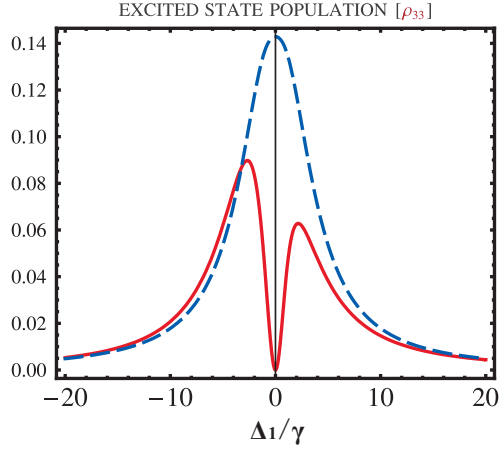}\\
  \caption{Dark-state condition. $g=G=\Omega=\gamma,\ \gamma_1=\gamma_2=\gamma, \Delta_2=0, \kappa=0
  $. The excited state population changes from high occupation ($\phi=\pi/2$, dashed blue
  line) to zero in dark-state $\phi=0$,(solid red line).}\label{phi332}
\end{figure}

Around $\phi=0,\pi$, the dark-state is very sensitive to the shift
of $\phi$. One resonance case neglecting ground state dephasing, the
excited state population can be solved as
\begin{equation}\label{c3r2}
    \rho_{33}=\frac{1}{3}\left(1-\frac{A}{A+6g^2\Omega^2\sin^2{\phi}}\right),
\end{equation}
where $A=2(g^2-\Omega^2)^2+g^2\Omega^2+8\gamma^2\Omega^2>0$ and
$G=g$ for dark-states requirement. It is obvious that when
$\phi=\pi/2,3\pi/2$, $\rho_{33}$ reaches the maximum while
$\phi=0,\pi$ for the minimum, which is the dark state with
$\rho_{33}=0$.

\begin{figure}
\centering
  \includegraphics[width=6.5 cm]{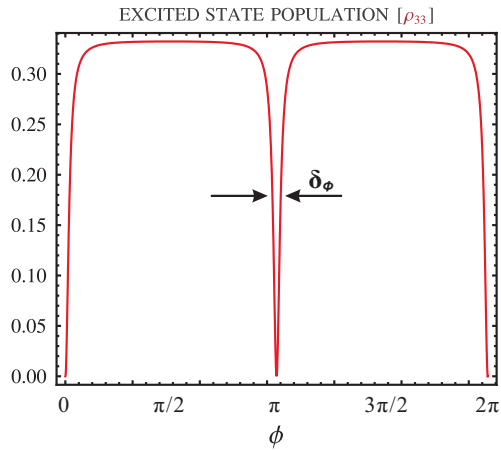}\\
  \caption{Phase sensitivity of excited state population. $g=G=\Omega=20\gamma,\
  \gamma_1=\gamma_2=\gamma, \Delta_1=\Delta_2=0, \kappa=0$. $\rho_{33}$ is extremely sensitive to
  the phase around dark-state condition and not sensitive to the phase in any other conditions.}\label{phi33}
\end{figure}

So full width at half maximum of the population versus $\phi$, as
illustrated in \Fref{phi33}, is
\begin{equation}\label{FMHW}
 \delta_\phi=2\sin^{-1}\sqrt{\frac{1}{6}+\frac{4\gamma^2}{3g^2}+\frac{(g^2-\Omega^2)^2}{3g^2\Omega^2}}.
\end{equation}
Thus for strong field coupling the atom states, the excited state
population will be very sensitive to the phase of microwave field
and so does the dark-state. On the contrary, the population is
extremely \emph{insensitive} to the phase of the microwave field if
\eref{c1} or \eref{c2} is not satisfied. This phase dependence
effect is similar to the phase sensitivity in \cite{DV}.

\section{EIT without Dark-States}
Dark state plays very important role in EIT, and total transparency
can be achieved only in dark-state. Up to now only $\Lambda$-type
atom can satisfy this condition \cite{RMPMF}. The other three-level
systems can exhibit EIT transparency window but could not generate
dark state \cite{RMPMF, 1995Fulton}, so that the transparency
windows are always accompanied with absorption. In our scheme, as
mentioned above, existence of the ground states dephasing limits the
applications of the dark-states irrespective the microwave field
exist or not. But with the help of the microwave field, the total
transparency can be achieved even with excited state occupied. A
previous study has already give an example \cite{sun}.

For $G\ll g$ as EIT requires, let $g=\zeta G$ where $\zeta\ll 1$.
The imaginary part of $\rho_{31}$ which represent the absorption of
the medium can be solved from the density matrix equation in
resonance case as
\begin{equation}\label{rho31}
    \mathrm{Im}[\rho_{31}]=\frac{\Omega\zeta\sin{\phi}+\kappa\zeta^2+\Or(\zeta^3)}{2g+\Or(\zeta^2)}
    \approx\frac{\Omega\zeta\sin{\phi}+\kappa\zeta^2}{2g}.
\end{equation}
Hence, if $\sin\phi=-\kappa \zeta/\Omega=-\kappa g/\Omega G$,
transparency happens ($\mathrm{Im}[\rho_{31}]=0$). Obviously this
condition is far from the dark-state condition and the excited state
is occupied.

\begin{figure}
\centering
  \includegraphics[width=7 cm]{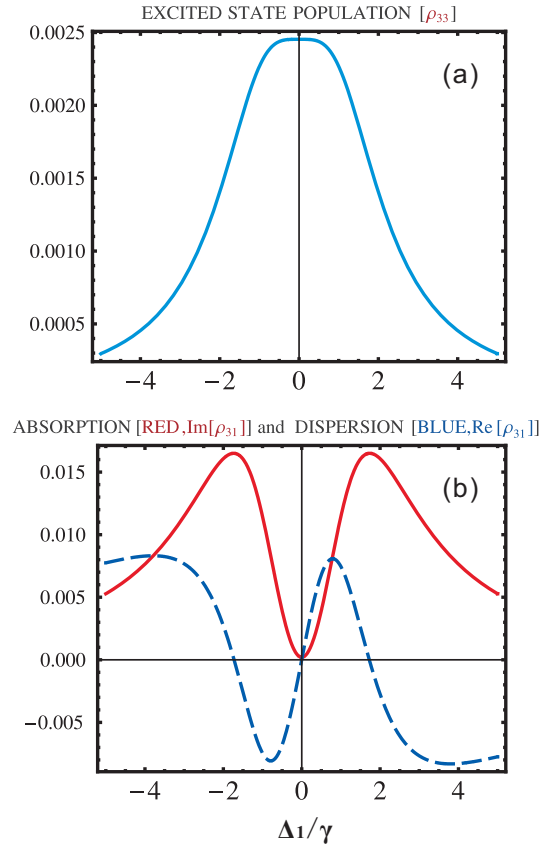}\\
  \caption{Transparency without dark-state. $g=0.05\gamma,\
  G=\gamma,\ \Omega=0.05\gamma,\ \gamma_1=\gamma_2=\gamma, \Delta_2=0, \kappa=\gamma $ and $\phi=-\pi/2$.
   a) The excited-state population. b) Absorption (red solid line) and dispersion (blue dashed line) properties. The transparency occurs with the excited-state population at its maximum.}\label{nondark}
\end{figure}

\begin{figure}
\centering
  \includegraphics[width=6.5 cm]{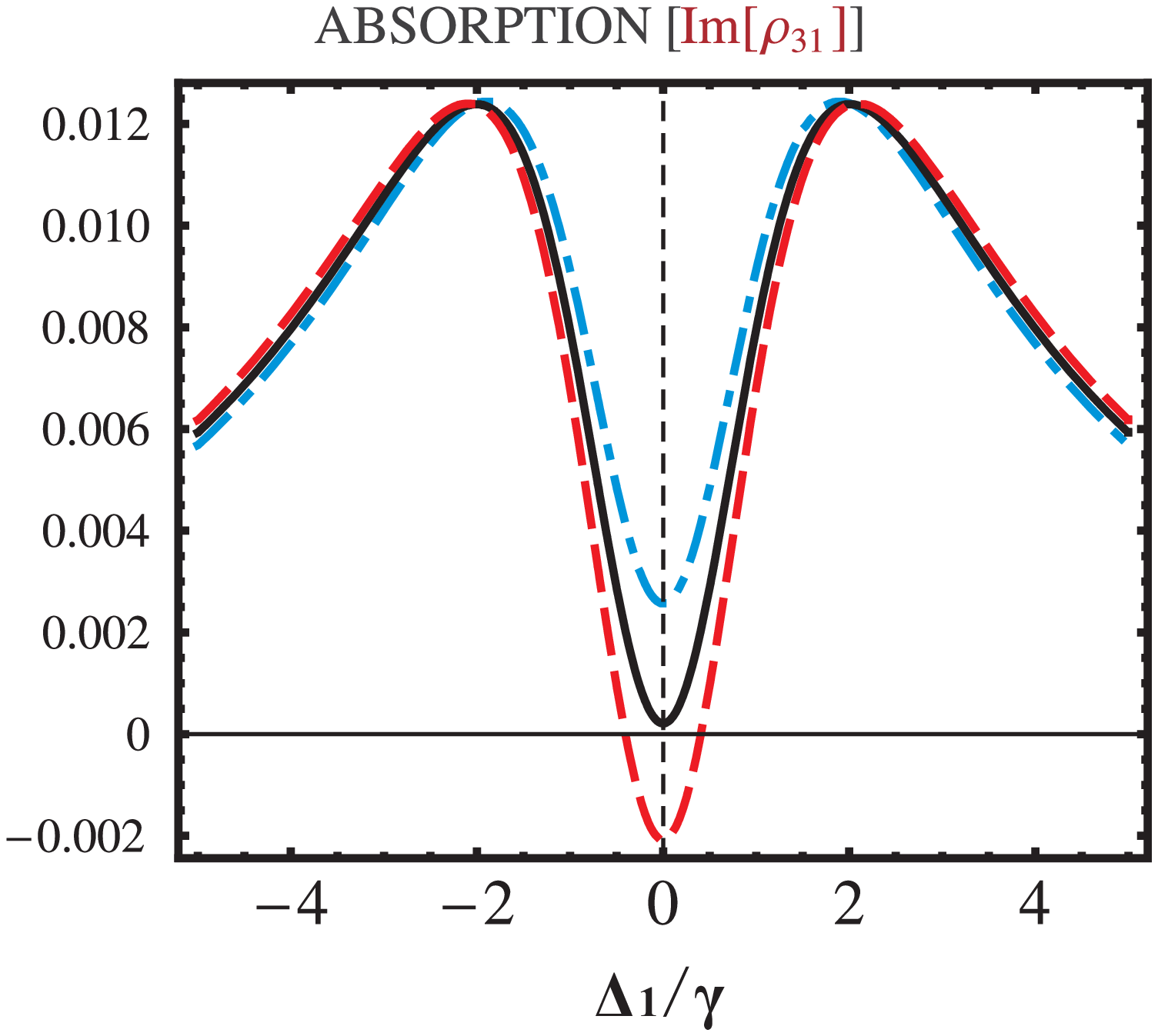}\\
  \caption{Absorption profile with different $\Omega$. $g=0.05\gamma,\
  G=\gamma,\ \gamma_1=\gamma_2=\gamma, \Delta_2=0, \kappa=\gamma $ and $\phi=-\pi/2$. 1) $\Omega=0.04$, absorption (dashed blue line);
  2) $\Omega=0.05$, transparency with neither loss nor gain (solid black line);
  3) $\Omega=0.06$, gain (dot dashed red line).}\label{difo}
\end{figure}

\begin{figure}
\centering
  \includegraphics[width=6.5 cm]{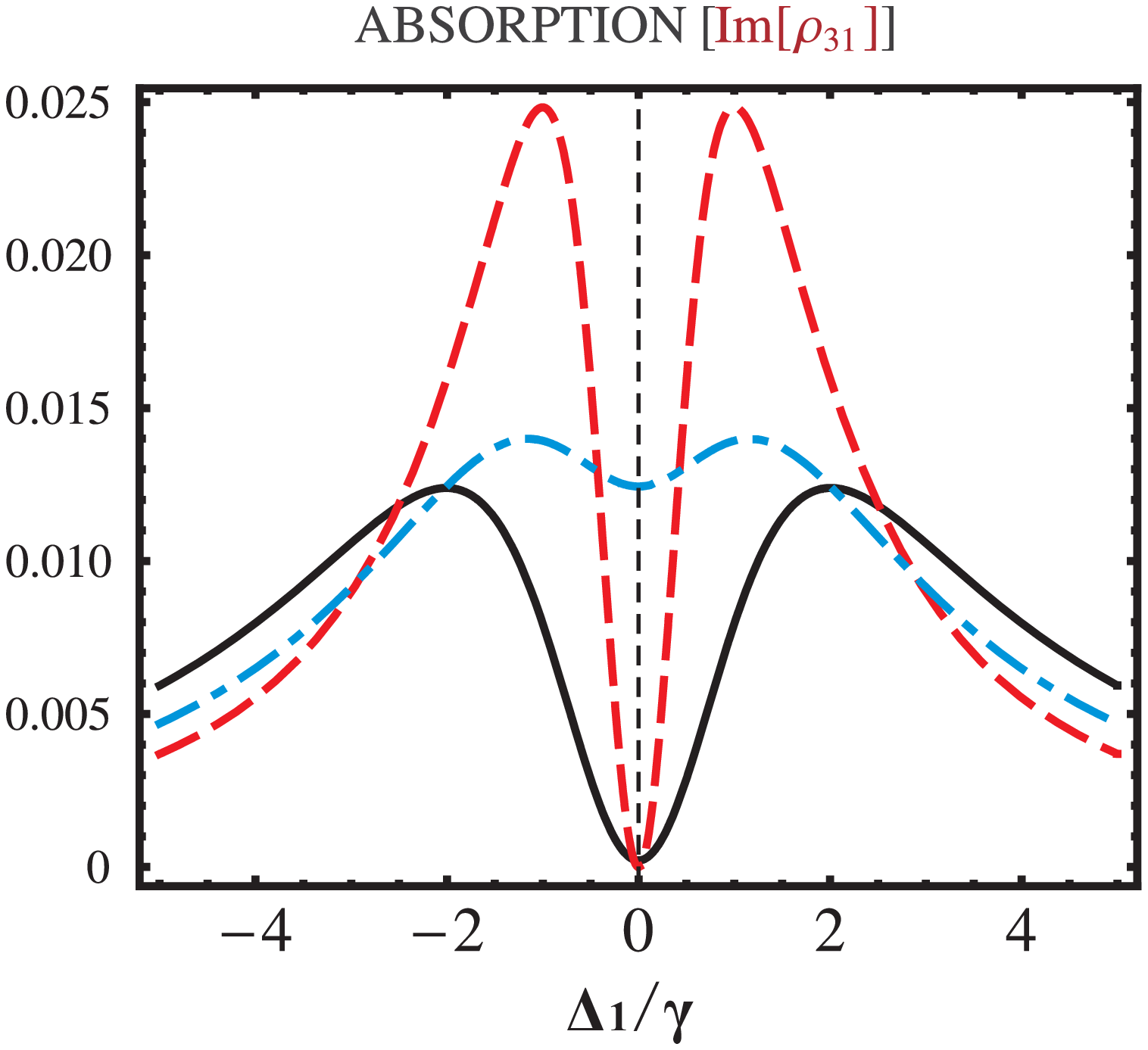}\\
  \caption{Comparison of absorption profile with or without microwave field. $g=0.05\gamma,\
  G=\gamma,\ \gamma_1=\gamma_2=\gamma, \Delta_2=0$.
  1) EIT with $\kappa=0$ and $\Omega=0$ (red dashed line); 2) EIT with $\kappa\gamma$ and $\Omega=0$ (blue dot dashed line), the transparency window is almost closed;
  3)EIT with $\kappa=\gamma$ and $\Omega=-\pi/2$ (black solid line), the transparency window is regained with the presence of the microwave field. }\label{difeit}
\end{figure}

\Fref{nondark} shows the excited state population and
absorption/dispersion properties. EIT does occur without
dark-states. This is because the presence of the microwave field can
generate gain rather than absorption. It is clear in \eref{rho31}
that $\Omega\sin{\phi}\zeta$ is the contribution from the microwave
field (gain if $\sin{\phi}<0$) and $\kappa\zeta^2$ is the absorption
caused by dephasing. Thus the transparency without dark-states can
be understood as the result of the balance between the absorption
(due to dephasing) and gain (due to microwave field). As in
\Fref{difo}, with different Rabi frequencies of the microwave field
the medium represents absorption, transparency and gain.
\Fref{difeit} shows the advantage of the microwave controlled EIT.
Without ground states dephasing, the medium is transparent to the
probe. If the ground states dephasing is large enough, however, the
transparency will disappear. Including the microwave field with
proper phase, the transparency is regained duo to the balance
between the microwave caused gain and the dephasing caused
absorption. Based on this method good EIT signal can be achieved
even in high dephasing medium.

\section{Conclusion}
In conclusion, we study a $\Lambda$-type EIT medium controlled by a
microwave field coupling the ground states. From Hamiltonian in
rotating frame the dark-state conditions are derived. Dark-states
under these conditions are the direct consequence of the coherence
among optical fields and microwave field. This kind of dark-state is
very sensitive to the relative phase between the microwave field and
the optical fields. However, with significant ground states
dephasing, the dark-states can never be achieved, which is similar
to EIT without microwave field. Even without dark-states, owing to
the presence of the microwave field, a gain mechanism is introduced
and the balance of the gain and the loss from the dephasing can be
balanced so that transparency can be regained without dark-states.
Thus, the microwave field provides more possible methods for quantum
control of the inner states of the atom structure, enhances EIT in a
high dephasing medium and benefits other EIT based phenomena.

\ack
This work is supported by the Key Project of the National
Natural Science Foundation of China (Grant No. 60837004) and Open
Fund of Key Laboratory of Optical Communication and Lightwave
Technologies£¬ (Beijing University of Posts and Telecommunications),
Ministry of Education, P. R. China.

\appendix
\section*{Appendix}
The full solution of the density matrix equation is very complicated
and not necessary to be given here. If $\rho_{33}=0$, the equation
gives the requirement for $\phi$ as
\begin{eqnarray}
\cos{\phi}=A\pm\sqrt{B},\nonumber
\end{eqnarray}
where
\begin{eqnarray}\label{A}
A=\frac{2 \gamma}{gG\Omega(4\gamma+\kappa)^2}\left\{\left[2 \gamma
g^2+\kappa  \Omega ^2-G^2 (2 \gamma +\kappa
   )\right] \Delta _1\right.\nonumber\\ \left.+\left[-(2 \gamma +\kappa ) g^2+\kappa  \Omega ^2+2 G^2
   \gamma \right] \Delta _2\right\},\nonumber
\end{eqnarray}
\begin{eqnarray}\label{B}
    B=\frac{\left[g^2 \kappa  G^2+2 \left(g^2+G^2\right) \gamma  \Omega
    ^2\right]
   \left[\kappa  \Omega ^2+2 \gamma  \left(g^2+G^2+2 \gamma  \kappa
   \right)\right]}{g^2 G^2 (4 \gamma +\kappa )^2 \Omega
   ^2}\nonumber\\+\frac{2 \gamma \left[\kappa  \Omega ^2+2 \left(g^2+G^2\right) \gamma \right]}{g^2
   G^2 (4 \gamma +\kappa )^4 \Omega ^2}\left\{g^2\left[2 \gamma  \left(\Delta _1-\Delta _2\right)-\kappa  \Delta
   _2\right]{}^2\right.\nonumber\\ \left.+G^2\left[\kappa  \Delta _1+2 \gamma  \left(\Delta _1-\Delta
   _2\right)\right]{}^2+2 \gamma \kappa \Omega ^2 \left(\Delta _1+\Delta
   _2\right){}^2\right\}.\nonumber
\end{eqnarray}
So $B>0$. Since
\begin{eqnarray}
B-(1-A)^2=\nonumber\\ \frac{4 \gamma ^2}{g^2 G^2 (4 \gamma +\kappa
)^4 \Omega
   ^2}\left[\left(g^2-G^2\right) \Omega +gG\left(\Delta _1-\Delta
   _2\right)\right]{}^2\nonumber\\+\frac{2 \gamma  \kappa }{g^2 G^2 (4 \gamma +\kappa )^4 \Omega ^2}\left\{2 g^2 \gamma  \kappa  G^2+4 \left(g^2+G^2\right) \gamma ^2 \Omega ^2\right.\nonumber\\ \left.+\left[g
   \left(G^2-\Omega ^2\right)-G \Omega  \Delta _1\right]^2+\left[G
   \left(g^2-\Omega ^2\right)-g \Omega  \Delta
   _2\right]^2\right\},\nonumber
\end{eqnarray}
it is clear that $B-(1-A)^2\geq 0$ and so $A+\sqrt{B}\geq1$ and
$A-\sqrt{B}\leq1$, the equality holds when $\left(g^2-G^2\right)
\Omega +gG\left(\Delta _1-\Delta_2\right)=0$ and $\kappa=0$.
Following the same step, one has $B-(1+A)^2\geq 0$ and so
$A-\sqrt{B}\leq -1$ and $A+\sqrt{B}\geq -1$, the equility holds when
$\left(g^2-G^2\right) \Omega -gG\left(\Delta _1-\Delta_2\right)=0$
and $\kappa=0$. The combination gives
\begin{eqnarray}
  A+\sqrt{B}\geq1, ``=" \ \mathrm{for}\ \phi=0,\kappa=0,\nonumber\\ \rightarrow\left(g^2-G^2\right)
\Omega +gG\left(\Delta _1-\Delta_2\right)=0; \nonumber\\
  A-\sqrt{B}\leq-1, ``=" \ \mathrm{for}\ \phi=\pi,\kappa=0, \nonumber\\ \rightarrow\left(g^2-G^2\right)
\Omega -gG\left(\Delta _1-\Delta_2\right)=0, \nonumber
\end{eqnarray}
which is exactly the same as \eref{c1} and \eref{c2}.


\section*{References}
\begin{thebibliography}{<num>}

\bibitem{Harris}Harris S E 1997 \emph{Phys. Toda}y \textbf{50} 36

\bibitem{RMPMF} Fleischhauer M, Imamoglu A and Marangos J P 2005
\RMP \textbf{77} 633

\bibitem{Hau} Hau L V, Harris S E, Dutton Z and Behroozi C H 1999 \emph{Nature} \textbf{397} 594

\bibitem{Akulshin1998} Akulshin A M, Barreiro S and Lezama A 1998 \PR A \textbf{57} 2996

\bibitem{hemmer1995} Hemmer P R, Katz D P, Donoghue J, Cronin-Golomb M, Shahriar M S and Kumar P 1995 \emph{Opt. Lett.} \textbf{20} 982

\bibitem{MXiao1} Li Y and Xiao M 1996 \emph{Opt. Lett}. \textbf{21} 1064

\bibitem{scully1999} Kash M M, Sautenkov V A, Zibrov A S, Hollberg L, Welch G R, Lukin M D, Rostovtsev Y, Fry E S and
Scully M O 1999 \PRL \textbf{82} 5229

\bibitem{harris2004} Braje D A, Bali V, Goda S, Yin G Y and
Harris S E 2004 \PRL \textbf{93} 183601

\nonum Bali V, Braje D A, Kolchin P, Yin G Y and Harris S E 2005
\PRL \textbf{94} 183601

\bibitem{Mxiao} Zhang Y, Anderson B and Xiao M 2008 \PR A \textbf{77} 061801(R)

\bibitem{MFandL} Fleischhauer M and Lukin M D 2000 \PRL \textbf{84} 5094

\bibitem{MFandL2} Fleischhauer M and Lukin M D 2001 \PR A \textbf{64} 022314

\bibitem{Bergman} Bergman K, Theuer H and Shore B W 1998 \RMP \textbf{70} 1003

\bibitem{DLCZ} Duan L M, ~Lukin M D, Cirac J I and Zoller P 2001 \emph{Nature} \textbf{414} 413

\bibitem{Lukinns} Eisaman M D, Andr\'{e} A, Massou F, Fleischhauer M, Zibrov A S and Lukin M D 2005 \emph{Nature} \textbf{438} 837

\bibitem{kuzmich} Chaneli\`{e}re T, Matsukevich D N, Jenkins S D, Lan S Y, Kennedy T A B and Kuzimich A 2005 \emph{Nature} \textbf{438} 833

\bibitem{DV} Kosachiov D V, Matisov B G, and Rozhdestvensky Y V 1992 \jpb \textbf{25} 2473

\bibitem{wilson} Wilson E A, Manson N B, Wei C and Yang L 2005 \PR A \textbf{72} 063813

\nonum Wilson E A, Manson N B and Wei C 2005 \PR A \textbf{72}
063814

\bibitem{shahriar} Shahriar M S and Hemmer P R 1990 \PRL \textbf{65} 1865

\bibitem{gemma}Yamamota K, Ichimura K and Gemma N 1998 \PR A \textbf{58} 2460

\bibitem{sun} Sun H, Guo H, Bai Y F, Han D A, Fan S L and Chen X Z 2005 \PR A \textbf{335} 68

\bibitem{agarwal} Agarwal G S, Dey T N and Menon S 2001 \PR A \textbf{64} 053809

\bibitem{arx} Li H, Sautenkov V A, Rostovtsev Y V, Welch G R, Hemmer P R and Scully M O 2009 arxiv:0903.1457v1

\bibitem{1995Fulton} Fulton D J, Shepherd S, Moseley R R, Sinclair B D and Dunn M H 1995 \PR A \textbf{52} 2302

\end{thebibliography}
\end{document}